\documentclass[sigconf,nonacm]{acmart}
\usepackage{amsfonts}
\usepackage{algorithmic}
\usepackage{textcase}
\usepackage{url}
\usepackage{textcomp}
\usepackage{multirow}
\usepackage{makecell}
\usepackage{xcolor}
\usepackage{tabularx}
\AtBeginDocument{%
  }

\setcopyright{none}

\begin{document}

\title{AI-Driven Intrusion Detection Systems (IDS) on the ROAD Dataset: A Comparative Analysis for Automotive Controller Area Network (CAN)}

\author{Lorenzo Guerra}
\orcid{0009-0008-6991-136X}
\affiliation{
  \institution{LTCI, Télécom Paris, Institut Polytechnique de Paris}
  \city{Palaiseau}
  \country{France}
}
\email{lorenzo.guerra@telecom-paris.fr}
\affiliation{
  \institution{Ampere Software Technology}
  \city{Guyancourt}
  \country{France}
}
\email{lorenzo.guerra@ampere.cars}

\author{Linhan Xu}
\orcid{0009-0002-3998-7060}
\affiliation{
  \institution{LTCI, Télécom Paris, Institut Polytechnique de Paris}
  \city{Palaiseau}
  \country{France}
}
\email{linhan.xu@utt.fr}

\author{Paolo Bellavista}
\orcid{0000-0003-0992-7948}
\affiliation{
  \institution{University of Bologna}
  \city{Bologna}
  \country{Italy}
}
\email{paolo.bellavista@unibo.it}

\author{Thomas Chapuis}
\orcid{0009-0005-6083-9080}
\affiliation{
  \institution{Ampere Software Technology}
  \city{Guyancourt}
  \country{France}
}
\email{thomas.chapuis@ampere.cars}

\author{Guillaume Duc}
\orcid{0000-0002-4804-3742}
\affiliation{
  \institution{LTCI, Télécom Paris, Institut Polytechnique de Paris}
  \city{Palaiseau}
  \country{France}
}
\email{guillaume.duc@telecom-paris.fr}

\author{Pavlo Mozharovskyi}
\orcid{0000-0002-1925-3337}
\affiliation{
  \institution{LTCI, Télécom Paris, Institut Polytechnique de Paris}
  \city{Palaiseau}
  \country{France}
}
\email{pavlo.mozharovskyi@telecom-paris.fr}

\author{Van-Tam Nguyen}
\orcid{0000-0003-3884-9520}
\affiliation{
  \institution{LTCI, Télécom Paris, Institut Polytechnique de Paris}
  \city{Palaiseau}
  \country{France}
}
\email{van-tam.nguyen@telecom-paris.fr}

\renewcommand{\shortauthors}{Guerra et al.}
\renewcommand{\shorttitle}{AI-Driven Intrusion Detection Systems (IDS) on the ROAD Dataset}

\begin{abstract}
The integration of digital devices in modern vehicles has revolutionized automotive technology, enhancing safety and the overall driving experience. The Controller Area Network (CAN) bus is a central system for managing in-vehicle communication between the electronic control units (ECUs). However, the CAN protocol poses security challenges due to inherent vulnerabilities, lacking encryption and authentication, which, combined with an expanding attack surface, necessitates robust security measures. In response to this challenge, numerous Intrusion Detection Systems (IDS) have been developed and deployed. Nonetheless, an open, comprehensive, and realistic dataset to test the effectiveness of such IDSs remains absent in the existing literature. This paper addresses this gap by considering the latest ROAD dataset, containing stealthy and sophisticated injections. The methodology involves dataset labelling and the implementation of both state-of-the-art deep learning models and traditional machine learning models to show the discrepancy in performance between the datasets most commonly used in the literature and the ROAD dataset, a more realistic alternative.
\end{abstract}

\keywords{AIoT, Intrusion Detection System, IDS, Controller Area Network, CAN, ROAD Dataset}

\maketitle

\section{Introduction}
With the rapid development of Internet of Things (IoT) and Internet of Vehicles (IoV) technologies, network-managed vehicles, such as Autonomous Vehicles and Connected Vehicles, are progressively taking the place of traditional automobiles. Within IoV systems, there are typically two core components: intravehicle networks (IVNs) and external networks. In IVNs, the core infrastructure is the Controller Area Network (CAN) bus, which enables communication among Electronic Control Units (ECUs).
In contrast, external vehicular networks utilize Vehicle-to-Everything technology to establish links between intelligent vehicles and other IoV elements.

The surge in connectivity has created a breeding ground for vulnerabilities. Malicious actions can be injected remotely through wireless communication systems, or physically through vectors like OBD-II ports, USB interfaces, CD drives, and more. In response to the growing threats, Intrusion Detection Systems (IDS) have emerged as a crucial defense mechanism. Leveraging the outstanding performance demonstrated by artificial intelligence, we aim to showcase the capabilities of various deep learning and conventional machine learning techniques in identifying anomalous behaviors.

By taking into consideration that the effectiveness and robustness of an Intrusion Detection System (IDS) are not solely dependent on the capabilities of the AI model in use, but also on the characteristics of the training dataset, we initiate the process of training different models using a unique dataset \cite{verma2022addressing} that includes 11 discrete categories of stealthy attacks, validated on a real vehicle, marking a significant departure from previous datasets, which were predominantly composed of few unrealistic attacks \cite{lampe2023cantrainandtest}. Starting by labeling \cite{verma2022addressing}, this paper's contributions extend to the evaluation of a wide range of supervised deep learning techniques, including the Transformer-based Attention Network (TAN) developed by Nguyen et al. (2023) \cite{10141599}, the Deep Convolutional Neural Network (DCNN) developed by Song et al. (2020) \cite{SONG2020100198}, and the model based on Long Short-Term Memory (LSTM) developed by Hossain et al. (2020) \cite{Hossain2020LSTMBasedID}, alongside various machine learning methods such as LightGBM, Random Forest (RF) and the LCCDE model \cite{lccde} applied to the aforementioned dataset. In conclusion, we conduct a comparative analysis, employing additional established datasets such as the HCRL Car Hacking dataset and the In-Vehicle Network Intrusion Detection Challenge dataset, eventually suggesting improvements.

The remainder of this paper is organized as follows. In Section \ref{sec:background}, we give a general background on the CAN protocol, its vulnerabilities and we review the literature on the topic. Section \ref{sec:datasets_description} provides an overview of the datasets used, with a specific focus on ROAD and its attacks. Subsequently, in Section \ref{sec:fundamentals}, we offer a concise depiction of the employed IDS techniques, followed by the presentation and analysis of the experimental setup and results in Section \ref{sec:environment_and_results}. To conclude, Section \ref{sec:conclusion} provides a summary of our findings.

\section{Background and literature review}\label{sec:background}
\subsection{Controller Area Network Bus}
\subsubsection{Protocol}
\begin{figure}
	\includegraphics[width=\linewidth]{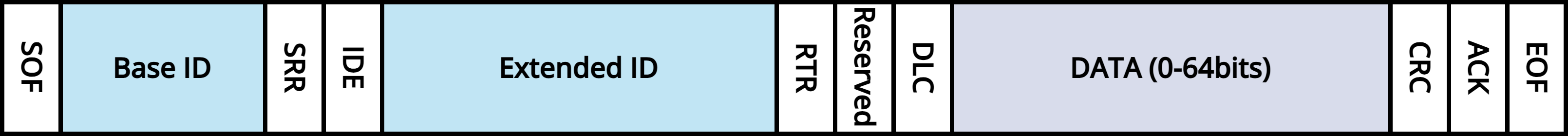}
	\caption{CAN bus data frame.}
	\label{fig:can-data-frame}
    \Description[CAN bus data frame.]{The picture shows an example of CAN bus data frame, with initial SOF, BaseID, IDE, Extended ID, RTR, Reserved, DLC, DATA from 0 to 64 bits, CRC, ACK, EOF}
\end{figure}

The CAN Bus, developed by Bosch in the 1980s for Local Area Networks (LAN), is a message-based protocol designed to enable numerous electric components, e.g., microcontrollers, ECUs, sensors, devices, and actuators, throughout the in-vehicle system to communicate with each other through a dual-wire bus. A standard CAN data frame (or message), featured in Figure~\ref{fig:can-data-frame}, consists of several fields. The most important fields are described as follows:
\begin{itemize}
\item Identifier (CAN ID): The CAN ID is represented by 11 bits in the standard format (CAN 2.0A) or 29 bits if the extended format is used (CAN 2.0B). It determines the priority of each message within the network. The CAN physical layer uses wired-AND signalling, meaning that a dominant level would be represented by a logical \texttt{0} and a recessive level by a logical \texttt{1}, so the lower the ID, the highest the priority.
\item Data Length Code (DLC): Located within the control field, this parameter indicates the number of bytes present in the data field, with a valid range extending from 0 to 8 bytes.
\item Data field: Contains the actual contents of the message, which can be up to 8 bytes in size. 
\item Cyclic Redundancy Check (CRC): a built-in error detection mechanism that ensures the reliability of the communication by detecting transmission errors and requesting a re-transmission if necessary.

\end{itemize}
\subsubsection{CAN vulnerabilities}

Because of its simplicity, the CAN bus lacks essential security measures:
\begin{itemize}
\item \textit{Missing Authentication}: ECUs do not have to authenticate to be able to transmit messages on the bus, every device can connect and it is immediately able to receive and transmit data, which makes the bus very flexible but vulnerable to intruders. In a normal scenario, the ECU corresponding to a precise functionality, for example, the wheel speed sensor communicating the speed at which the vehicle is going, will only send information concerning its domain, and it would never transmit frames reporting the coolant temperature. But protocol-wise nothing prevents the ECU from sending any type of information, which in case of an intruder could lead to spoofing attacks performed with ease.
\item \textit{Missing Encryption}: the data transmitted on the CAN bus is transparent and it is broadcasted to every ECU, meaning that any attacker could eavesdrop on unencrypted communication without anyone noticing. Combined with the missing authentication, this can lead to more complex attacks harder to detect since the normal behavior of ECUs can be anticipated, potentially suppressing the real messages. Spoofing attacks are also easy to perform by exploiting the transparency and the missing authentication, making detection exceptionally hard if such attacks are performed correctly.
\end{itemize}
Some approaches to secure the communication on the bus have been experimented during the years \cite{vector-encryption}, but unfortunately they have not gained enough traction.
Consequently, the protocol does not respect 2 of the 3 fundamental properties focused on by information security principles, which are summarized by the CIA triad, \textit{Confidentiality, Integrity, Availability}:
\begin{itemize}
    \item \textit{Confidentiality} is not respected because of the missing encryption, since any intruder is able to eavesdrop on all the traffic just by connecting to the CAN bus.
    \item \textit{Integrity} is respected thanks to the CRC, detecting bit errors, while error frames force the bus dominant longer than the bit stuffing rule. Furthermore, automatic re-transmission occurs until the error limit is reached.
    \item \textit{Availability} is not respected because the authentication is missing, anyone connected to the bus can flood it with high-priority frames, overwhelming the communication and preventing the ECUs from sending any message.
\end{itemize}

The most difficult task for an attacker, considering the critical vulnerabilities of the CAN protocol, would be to access the bus. The main attack vectors are two \cite{experimental-security-analysis-automobiles}, one is physical access, the other one is through one of the many wireless interfaces.

To reach the bus through physical access, the simplest way would be through the OBD-II port, mentioned in the previous section. This connector exposes the CAN bus without needing any authentication, so an intruder would be able to eavesdrop and inject messages just by connecting a malicious component to it. It is also possible to leave the component attached indefinitely, concealing it in the dashboard, where the OBD-II port is most commonly found. Another way to get physical access would be to install a counterfeit component in the car, so the supply chain itself needs to be secured, and only authorized repair shops should be able to replace components.
Reaching the CAN bus through one of the wireless interfaces is less simple and would require such interfaces to be themselves vulnerable. Because of the large number of these devices in recent vehicles, it is not rare to find a vulnerability in one of them leading to access to the CAN bus.

A challenge-response mechanism protects ECUs from being reflashed and reading sensible memory. This challenge-response is performed through a seed-to-key challenge, but both seed and key are fixed and stored in the device. The system allows only an attempt every 10 seconds, but due to the low complexity, it is sufficient to have access for 7 and a half days to be able to crack the challenge-response and therefore reflash every ECU \cite{experimental-security-analysis-automobiles}.

The CAN bus exposes also a diagnostic system called Device Control, a powerful tool that is meant to be used for testing purposes, and it can modify the internal state of the ECUs. Paired with the vulnerabilities already described, it is easily exploitable and it can lead to dangerous consequences. This service takes an argument called a Control Packet Identifier (CPID), which specifies a group of controls to override, and each one can take up to five bytes as parameters, specifying which controls in the group are being overridden, and how to override them \cite{experimental-security-analysis-automobiles}.

\subsection{Literature review and research gaps}

One of the first security analyses on real cars has been made in 2010 by Koscher et al. \cite{experimental-security-analysis-automobiles}, assessing the behavior of the vehicle during specific attacks. They leveraged the security vulnerabilities of CAN to control a wide range of critical functions, like stopping the engine, disabling the brakes, or even braking on-demand individual wheels.
However, they do not yet prove that these attacks are feasible remotely. In 2011 Checkoway et al. \cite{checkoway2011comprehensive} analyzed purely the external attack surface of one vehicle to assess if it might be vulnerable to remote attacks exploiting the wireless interfaces present in the car, and they found out that it is indeed possible through multiple short-range and long-range vectors like Bluetooth or cellular interfaces. The results of Checkoway et al. are particularly concerning if we take into consideration the paper published by Koscher et al.: these two studies show that intruders can preserve their anonymity controlling critical features of vehicles from long distances, potentially causing disastrous, life-threatening consequences.

Charlie Miller and Chris Valasek are two researchers working on automotive cybersecurity who gained a reputation during the years by publishing multiple relevant papers on the topic. In 2013 \cite{adventures-in-automotive-networks} they tinkered with the CAN bus, exploiting its vulnerabilities to take control of critical functionalities of two different cars; in 2014 \cite{survey-of-remote-sufaces} they took Checkoway et al.'s previous research further by identifying the remote attack surface for a large range of vehicles, identifying the ECUs' capabilities and estimating the difficulty of remote exploitation; in 2015 \cite{remote-exploitation-miller-valasek} they brought to conclusion 3 years of research disclosing a long-range remote attack that can be performed on Jeep Cherokees without any physical interaction, causing Fiat Chrysler to recall 1.4 million vehicles \cite{fca-recall}. One year later, in 2016 \cite{can-message-injection}, they published another paper explaining how they performed a masquerade attack on the Jeep Cherokee and how they were able to deal with message confliction. 

Charlie Miller and Chris Valasek believe that attack detection represents an inexpensive and accurate way to greatly improve the security of CAN networks \cite{survey-of-remote-sufaces}. Current research on the topic focuses extensively on machine learning \cite{bari2023intrusion} and deep learning \cite{lin2022using} approaches in order to reach an increasingly higher accuracy. Unfortunately, these systems are not yet perfect: in 2021 Bhatia et al. \cite{bhatia2021evading} explained how they developed a masquerade attack called DUET which could evade detection of Voltage-based Intrusion Detection Systems. This type of IDS is able to observe the voltage fingerprint of ECUs, thus recognizing the ones belonging to attackers. However, the DUET attack showed that VIDS are not a silver bullet against intrusions, so the authors introduced a lightweight defense mechanism called RAID, suggesting to modify the frame format in a protocol-compatible way in order to safeguard against the corruption of ECUs' voltage fingerprints.

According to its design, an Intrusion Detection System can be categorized into two major families: anomaly-based and signature-based detection methods \cite{lokman2019intrusion}. While signature-based IDSs compare the traffic with a database of known attack patterns, anomaly-based IDSs can be powered by machine learning techniques to model the normal behavior of the machine and they are able to detect unseen attacks as well. 

Although prior AI-based IDS models have yielded to promising results \cite{lccde}\cite{Hossain2020LSTMBasedID}\cite{SONG2020100198}\cite{10141599}, there exists a remarkable gap in testing them on a more realistic dataset including sophisticated attack patterns.
It is worth noting that the effectiveness of an IDS model depends heavily on the quality of the data it was trained on. Previously introduced approaches were tested on datasets characterized by a limited number of attack types, typically only three or four, which are considered simple and unrealistic as they are conducted in a testbed environment \cite{lampe2023cantrainandtest}. These attacks are easily detectable and not necessarily verified on a real car. To build a robust IDS able to detect stealthy attacks it is imperative to test the model against high fidelity attack scenarios.

\section{Datasets description}\label{sec:datasets_description}
\subsection{ROAD dataset and Labeling}\label{road-dataset-description}
The Real ORNL Automotive Dynamometer (ROAD) dataset \cite{verma2022addressing}
 includes 33 attack captures spanning 30 minutes and 12 ambient captures over 3 hours (Table~\ref{tab:road-num-logs}). It originates from a real undisclosed vehicle driven on a dynamometer of the Oak Ridge National Laboratory (ORNL) during 3 and a half hours of activity, reproducing various benign driving behaviors, while ambient and attack data were collected.
 
The logs are collected with the candump utility \cite{candump} and the output format is the following:

\begin{center}
$\underbrace{\mathtt{(1040000000.000682)}}_{Unix Timestamp}  \underbrace{\mathtt{can0}}_{Channel} \ \underbrace{\mathtt{0BA}}_{ID}\mathtt{\#}\underbrace{\mathtt{04B7EC04000602C8}}_{Data Field} $
\end{center}

\begin{table}[hb]
    \caption{Overview of the attacks contained in the ROAD dataset.}
    \begin{center}
    \begin{tabularx}{0.95\linewidth}{cr}
    \toprule
    \textbf{Attack Types} & \textbf{\shortstack[r]{\# Attack \\ Samples}}\\
    \midrule
    Accelerator Attack in Drive &-\\
    Accelerator Attack in Reverse &-\\
    Correlated Signal Fabrication Attack &5493\\
    Correlated Signal Masquerade Attack &5493\\
    Fuzzing Attack &1059\\
    Max Engine Coolant Temp Fabrication Attack&43\\
    Max Engine Coolant Temp Masquerade Attack&43\\
    Max Speedometer Fabrication Attack &11694\\
    Max Speedometer Masquerade Attack &11694\\
    Reverse Light Off Fabrication Attack &5480\\
    Reverse Light Off Masquerade Attack &5480\\
    Reverse Light On Fabrication Attack &8034\\
    Reverse Light On Masquerade Attack &8034\\
    \bottomrule
    \end{tabularx}
    \end{center}
    \label{tab:road-num-logs}
\end{table}

The dataset features eleven distinct classes depicted in Table~\ref{tab:road-num-logs} for which metadata for labelling was provided. An extra attack called "accelerator attack" is included in the dataset, although the labelling information was not disclosed by the authors due to the seriousness of the vulnerability.

The attack types can be summarized in 3 categories:
\begin{enumerate}
  \item \textit{Fuzzing Attack}: Frames with cycling IDs have been injected with a payload set to the maximum value (0xFFFFFFFF). This attack caused various physical effects, for example: the accelerator pedal became ineffective, the dash lights and headlights were illuminated and the seat positions moved.
  \item \textit{Targeted ID Fabrication Attacks}: These attacks are performed injecting a message immediately after the legitimate message targeting a specific ID. In this case only the target signal is modified. Differently from most of public datasets, the fabrication attacks in the ROAD dataset contain one single frame injected between ambient frames of the same ID, since only a single injected message is needed after each legitimate message for this attack to be successful. Because these are real, physically verified attacks with the minimally occurring injected frames (due to the flam delivery), they provide perhaps the best (i.e., most stealthy/most difficult to detect), current, public data for testing frequency-based IDSs \cite{verma2022addressing}.
  \item \textit{Masquerade Attacks}: These attacks are performed during post-processing by removing the legitimate target ID frames preceding each injected frame, making it appear as though only the spoofed messages are present during the injection interval. The paper explains that even though the effect on the vehicle was verified and the CAN protocol was not violated, this attack has not been performed directly on the vehicle because of the complexity of the implementation. There are no available public captures of real masquerade attacks. This attack can be defined as \textit{Timing Opaque (T.O.)} because it does not disrupt normal timing or ID distributions, so it is very unlikely that frequency-based approaches could provide accurate detection, while the previous types can be defined as \textit{Timing Transparent (T.T)} because a frequency-based approach would be able to detect them.
\end{enumerate}

The log files themselves were not labeled, but the authors provided a file with the capture metadata in JSON format that allowed us to assign the labels based on the conditions specified (i.e. the interval during which the attack occurred, the ID used during the injection and the injected bytes in the data field).

\subsection{The HCRL Car-Hacking dataset} 
The Car-Hacking dataset \cite{car-hacking-dataset}, according to Rajapaksha et al. (2023) \cite{rajapakshaetal}, is the most popular dataset in literature, and in many recent publications it seems to be used without citation \cite{verma2022addressing}. It is generated by injecting CAN messages into the Hyundai YF Sonata's CAN bus through the OBD-II port and includes both attack and regular messages. Each sample is composed by a timestamp, the CAN ID, the DLC and up to 8 bytes of the data field. The attack types are four:
\begin{itemize}
    \item \textit{DoS attack}: This attack is simply performed by transmitting messages on the CAN bus with the ID containing only dominant bits (0) (i.e. ID=\texttt{0x000}) every 0.3 milliseconds. As seen in the previous section, the CAN bus is using wired-AND signaling, which means that the messages with the ID containing the largest number of dominant bits will be prioritized over any other ECU, which will have to patiently wait for the attack frame to be completely transmitted before trying to send another one. Therefore, an attacker with access to the CAN bus would simply need to continuously inject messages using this technique and the correct functioning of the system would be disrupted, needing a manual restart of the car \cite{experimental-security-analysis-automobiles}.
    \item \textit{Fuzzy attack}: CAN messages with spoofed random ID and DATA values are injected every 0.5 milliseconds, causing an unexpected behavior of the vehicle.
    \item \textit{RPM/Gear attack}: CAN messages with spoofed ID relative to the RPM or Gear ECU are injected every 1 millisecond.
\end{itemize}
For each of the four attack datasets, the total number of messages is approximately 4,000,000 (Table \ref{tab:details-car-hacking}), of which 13\% to 16\% are attack messages, therefore maintaining a relatively balanced attack proportion. There is a fifth file containing normal data in log format, but unfortunately, most of the papers employing this dataset seem to exclude the data contained in this additional log file, likely due to its limited benefits, as the other four CSV files already provide a considerable amount of benign messages.

\begin{table}[ht]
    \caption{Overview of the Car-Hacking Dataset.}
    \centering
    \begin{tabularx}{0.98\linewidth}{crrr}
        \toprule
        \textbf{Name of the Attack} & \textbf{\shortstack[r]{\# Total\\ Messages}} & \textbf{\shortstack[r]{\# Normal\\Messages}} & \textbf{\shortstack[r]{\# Attack\\Messages}}\\
        \midrule
        DoS Attack&3,665,771&3,047,062&587,521\\
        Fuzzing Attack&3,838,860&3,259,177&491,847\\
        Gear Spoofing Attack&4,443,142&3,805,725&597,252\\
        RPM Spoofing Attack&4,621,702&3,925,329&654,897\\
        Normal Run Data&988,871&988,871&---\\
        \bottomrule
    \end{tabularx}
    \label{tab:details-car-hacking}
\end{table}

\begin{table*}[ht]
    \caption{Overview of In-Vehicle Network Intrusion Detection Challenge dataset.}
    \centering
    \begin{tabularx}{0.66\textwidth}{crrrrr}
        \toprule
        \textbf{\shortstack[l]{\# Name of the\\Vehicle}} & \textbf{\shortstack[r]{\# Total\\ Messages}} & \textbf{\shortstack[r]{\# Normal\\Messages}} & \textbf{\shortstack[r]{\# Flooding\\Messages}} & \textbf{\shortstack[r]{\# Fuzzy\\Messages}} & \textbf{\shortstack[r]{\# Malfunction\\Messages}}\\
        \midrule
        Chevrolet Spark&313,934&291,897&14,999&3,043&3,995\\
        Hyundai Sonata&388,173&353,783&17,093&9,095&8,202\\
        Kia Soul&569,516&527,061&16,072&21,613&4,770\\
        \bottomrule
    \end{tabularx}
    \label{tab:details-in-vehicle}
\end{table*}

\subsection{The In-Vehicle Network Intrusion Detection Challenge dataset}
The In-Vehicle Network Intrusion Detection Challenge dataset \cite{in-vehicle-dataset} was used during the "Information Security R\&D Data Challenge 2019" and includes traffic data collected from three vehicles in stationary state: a Huyndai Sonata, a Kia Soul, and a Chevrolet Spark. We have used the datasets collected during the preliminary round, since they include three attack categories (Flooding, Fuzzy, and Malfunction) compared to the two attack categories of the final round. The attack types are the same of the Car-Hacking dataset, but with different names. The dataset is labelled and, like the Car-Hacking dataset, each sample is represented by a timestamp, a CAN ID, the DLC and the data field. For all three types of vehicles, the proportion of "Flooding" attacks is relatively high, about 20\% of the total messages, compared to only about 8\% for the other two types (Table \ref{tab:details-in-vehicle}). The application of the same attack types and setup on three different vehicles makes this dataset unique and enables us to test transfer learning approaches on it, in order to improve the generalizability of the model.

\section{Fundamentals of machine learning and deep learning IDS Models}\label{sec:fundamentals}

\subsection{LCCDE and ML approaches}
To compare neural network approaches to traditional machine learning ones, gradient boosting and decision tree based approaches have also been tested on the available datasets.
Specifically we considered Random Forest, LightGBM and the LCCDE model. These methods, like other machine learning algorithms, are preferable because of their speed, efficiency and explainability. Furthermore, the Leader Class and Confidence Decision Ensemble (LCCDE)~\cite{lccde}, an ensemble model based on gradient boosting algorithms, was able to reach state-of-the-art results. 
The ensemble algorithm is divided into two stages: training and prediction. In the training stage, the three base models are trained on the labeled data. For each class, the leader is selected based on the best F1-score of the models, obtained from cross-validation. In case of a tie in F1-score, the model with the highest speed is chosen.

In the prediction step, every model is first used to make predictions. Decisions are then taken based on the predicted classes:
\begin{itemize}
    \item If the predicted classes are the same, then this class will be the final prediction.
    \item If the predicted classes are all different, the leader model associated with each predicted class is cross-referenced with the base learner that made the corresponding prediction. If a single pair of leader and base models align, their jointly predicted class is adopted as the final prediction. In cases where multiple pairs exist, the predicted class with the highest confidence level is selected as the ultimate prediction.
    \item If only two of the predicted classes match, then the respective leader model is chosen for the final prediction.
\end{itemize}
The traditional machine learning models and the LCCDE were fed single frames with ID, DLC (if available) and data field as input and all these approaches involve multiclass classification, allowing a more precise prevention system based on the specific attack type detected.

\subsection{LSTM-based model}
LSTMs are a type of Recurrent Neural Networks (RNNs) that are frequently and effectively applied for sequence classification tasks, therefore we have selected the state-of-the-art LSTM model described by Hossain et al.~\cite{Hossain2020LSTMBasedID} to test this approach on the previously specified datasets, since it achieved outstanding results on the Survival Analysis Dataset \cite{survival-analysis-dataset}. The authors have tested many hyperparameters for both binary and multiclass classification. The best performing model was composed by a single LSTM layer with 512 units using single time steps and batches of size 512 as input. Each input sample includes ID, DLC (if available) and the bytes of the data field. We have performed multiclass classification on every dataset and binary classification only on the ROAD dataset, to compare its results with the DCNN and the TAN models. However, the approach of analyzing single messages can be considered vulnerable to replay attacks and since the authors used single time steps, the utility of a LSTM for this purpose is questionable.

\begin{figure*}
    \centering
	\includegraphics[width=0.52\textwidth]{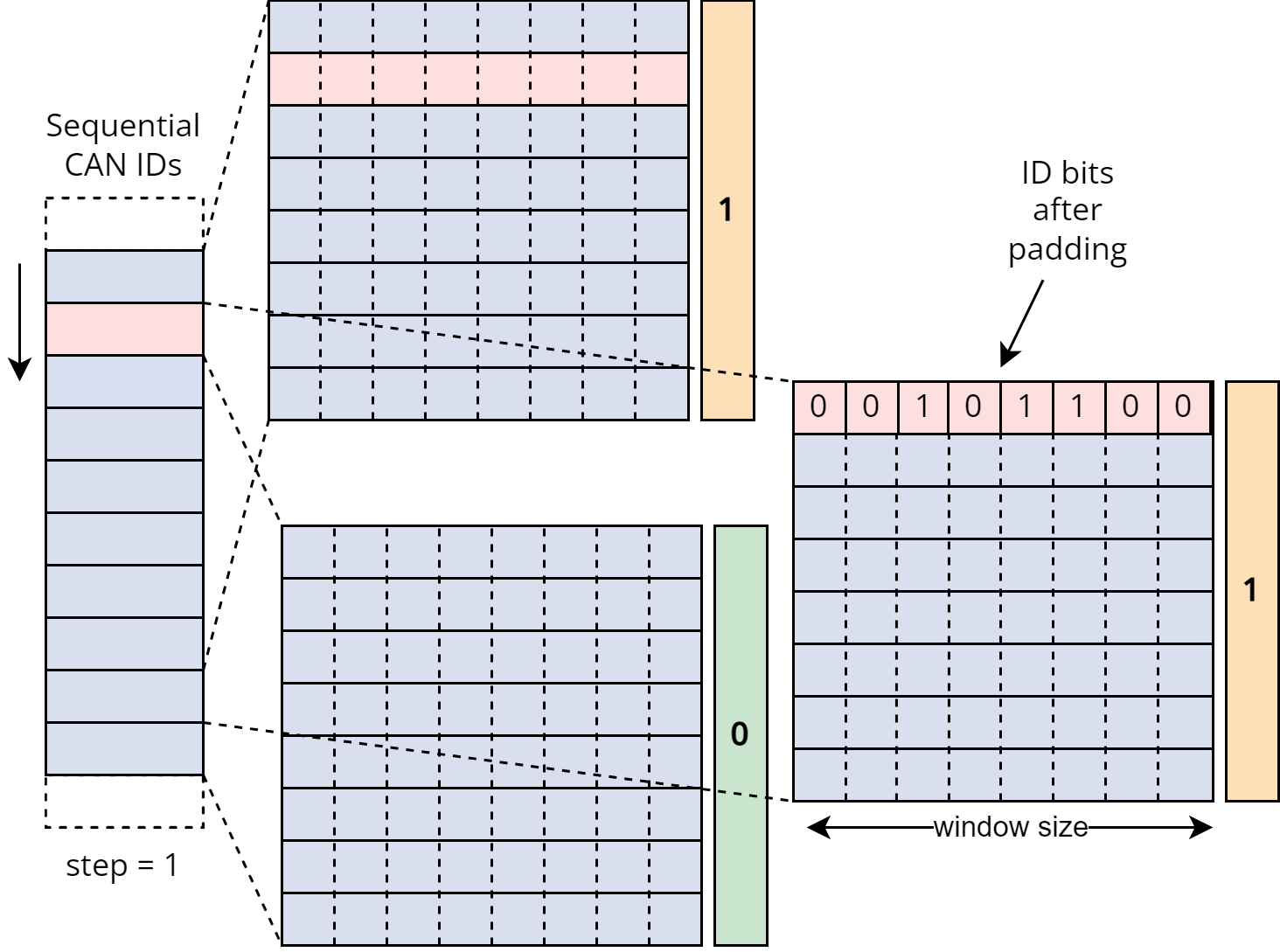}
	\caption{Frame creation process of the DCNNv2.}
	\label{fig:dcnn_frame_creation}
    \Description[Frame creation process of the DCNNv2]{The picture shows the improved process of how the grids are created and labeled: if an attack frame's ID is present in the sequence, the frame is labeled as an attack (1), otherwise it is labeled as normal (0). Horizontally, the bits of the ID are positioned in the grid. It also shows the step=1 to create the sequences.}
\end{figure*}

\subsection{DCNN}\label{subsection:DCNN}
The IDS model proposed by Song et al. \cite{SONG2020100198} is based on a convolutional neural network (CNN), aimed at detecting the patterns of IDs appearing in the CAN traffic. During the training phase, the model employs a frame builder to extract sequences of IDs from the logged traffic data. To facilitate CNN processing, which typically works with grid-formatted input, the temporal sequential patterns of CAN traffic are processed as spatially-local correlation. The frame builder extracts the 29-bit identifiers, which are the maximum number of bits for representing an ID when using the 'extended format'~\cite{can-specification}, from the most recent 29 messages and builds a $29\times29$ bitwise frame by stacking them. A sliding window of size 29, with a step of 29 is used. This means that attack patterns might extend across 2 different frames and could not be easily recognized by the DCNN, which is unable to retain memory across frames. For this reason, we tested a different approach modifying the step of the sliding window to one, enabling the model to catch every pattern involving IDs and extending for any 29 consecutive frames. The frame creation process is illustrated in Figure \ref{fig:dcnn_frame_creation}.

The frame builder uses pure bits as input data without preprocessing, ensuring computational efficiency, given the high volume of messages on a CAN bus. The bit representation of the identifier explicitly illustrates the fluctuation of identifier patterns. The bit representation of the identifier, denoted as:
\begin{equation}
    ID=b_i \ (for \ i=0, ..., 28)
\end{equation}
is assembled into a data frame as:
\begin{equation}
    FRAME=b_ij \ (for \ i=0, ..., 28)
\end{equation}
where \(b_{ij}\) represents the \(j\)-th bit value of the \(i\)-th ID in a frame. Each frame (grid) is then labeled as one if it contains an attack message, or zero for normal traffic.

The DCNN model itself is based on the Inception-ResNet model \cite{szegedy2016inceptionv4inceptionresnetimpactresidual}, which is originally designed to classify $299\times299\times3$ input image data into 1000 classes, but has been reduced by the authors to classify more efficiently $29\times29\times1$ into only 2 classes. Compared to the original model the proposed DCNN excludes the Inception-ResNet-C block, while keeping the stem, Inception-ResNet-A, -B, Reduction-A and -B, average pooling, dropout, and softmax blocks, although they have been internally simplified to adapt to the studied case. The final model uses only 2\% of the memory of the original one, and only the 18\% of the original parameters, making it more suitable for use in restricted environments, like automotive applications.

\subsection{TAN}
The last approach for intrusion detection that we adopted in this study relies on a transformer-based architecture developed by Nguyen et al. \cite{10141599}. Differently from the original transformer architecture \cite{vaswani2023attention}, the TAN architecture only uses the encoder, the most essential component for classification, thus resembling the highly influential ViT model produced by Google \cite{dosovitskiy2021image}. Furthermore, the model has been tested for transfer learning, which aims to address a significant challenge in in-vehicle intrusion detection. This method enables the application of a model trained on a large amount of data from a specific vehicle to a different vehicle for which less data is available.

\begin{figure*}
\centering
	\includegraphics[width=0.5\textwidth]{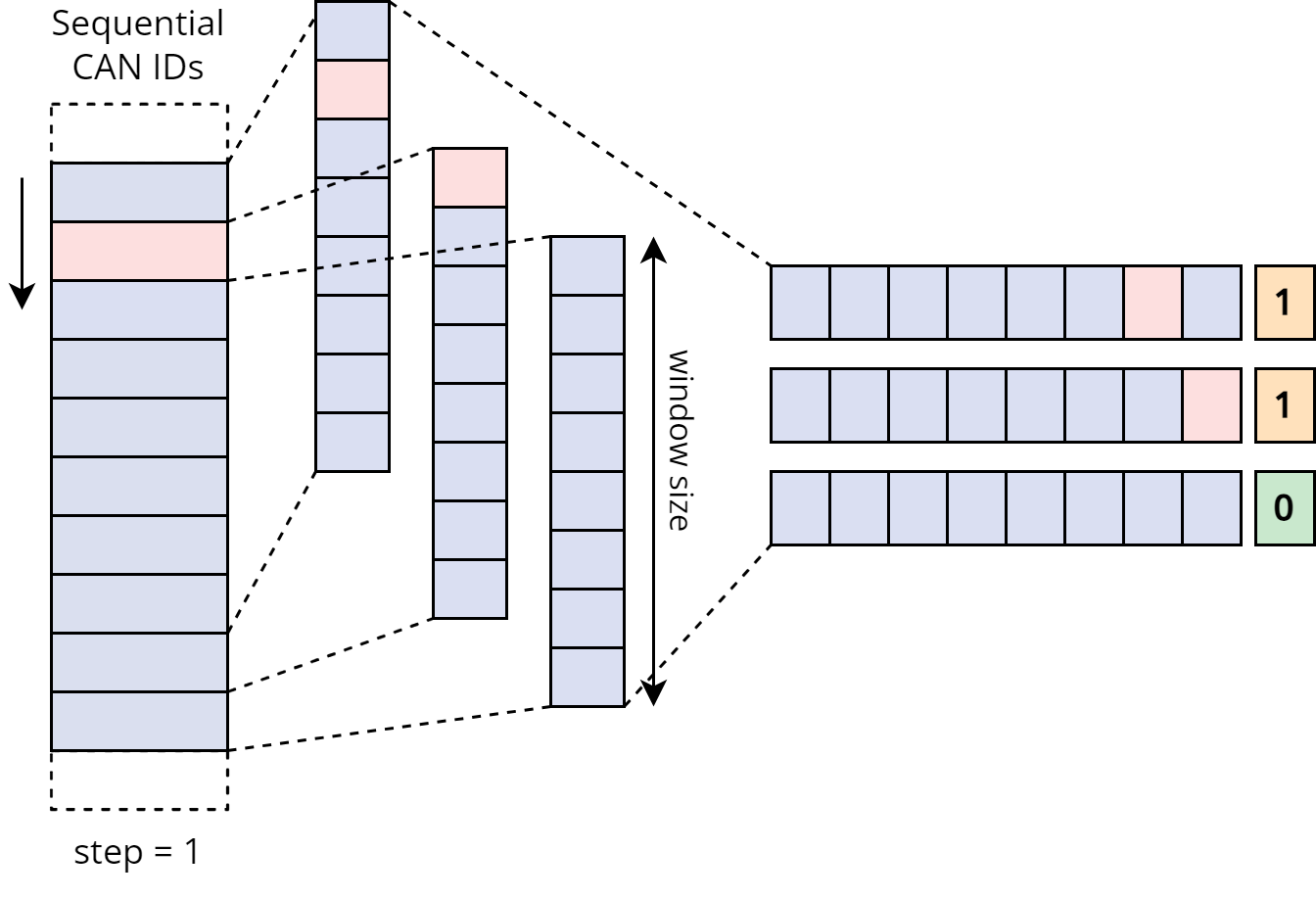}
	\caption{Sequence creation process for the TAN.}
	\label{fig:tan_sequence_creation}
    \Description[The picture shows the sequence creation process for the TAN model]{Similarly as for the DCNN, sequences with step=1 are created and if they contain an attack frame's ID, shown in red, then they are labelled as attack (1), 0 otherwise. }
\end{figure*}

The authors have explored two different approaches, one which analyses single messages and their every field, but it is vulnerable to replay attacks, and one that analyzes sequences of CAN IDs. We have chosen the latter as it is a more solid choice and it could achieve very good results, despite part of the information relevant to intrusion detection is discarded. The sequence creation process involves a sliding window with a step equal to one. Multiple window sizes have been tested and we have selected a window size of 16 because of its balance between performance and memory requirements. The sequence creation process is illustrated in Figure \ref{fig:tan_sequence_creation} and consists in grouping sequences of consecutive CAN IDs, which are labelled as 1 if they contain an attack message, or 0 otherwise. This choice implies the approach of binary classification, therefore impeding us to distinguish what types of attack are detected.

\section{Experimental environment and results analysis}\label{sec:environment_and_results}

\subsection{Comprehensive Framework: Experiment Setup, Data Preprocessing, and Performance Evaluation Metrics}

Our research involves multiple models and multiple datasets, which demand distinct preprocessing for each case. In the case of the ROAD dataset, the logs are not labelled, so we parsed the metadata made available by the authors and used it to systematically parse and label each sample. Finally, we merged every log for the multiclass classification task, while we separately merged each log of every attack type for binary classification, resembling the HCRL Car-Hacking dataset setup.

Commonly to every dataset, we parsed each field into CAN ID, DLC (unavailable for the ROAD dataset) and data field. We padded the latter to 8 bytes and then we divided it into 8 different features, allowing detection of attacks targeting specific bytes of this field. Each field is then converted from hexadecimal to decimal format, in order to be processed by the selected models, and every dataset is split into training and test sets with a 80/20 ratio.

It's worth noting that the ROAD dataset is highly imbalanced, as it can be seen on Table~\ref{tab:road-num-logs}, and some classes need to be oversampled. For this reason we employed the SMOTE algorithm \cite{Chawla_2002} on the training set to increase the number of samples of each attack to a fixed value of 100,000, therefore ensuring an equal distribution between them, that otherwise could have not been detected from any metric produced by our models after building the sequences. The In-Vehicle Network Intrusion Detection Challenge dataset has been oversampled in the same way.
The resulting data can be directly fed to the traditional machine learning models and the LSTM-based model, but it needs to undergo further processing in order to create the frames needed by the DCNN and the sequences needed by the TAN, built as described in Section \ref{sec:fundamentals}, in Figure \ref{fig:dcnn_frame_creation} and Figure \ref{fig:tan_sequence_creation}.

The Random Forest, LightGBM, LCCDE and LSTM-based model perform multiclass classification on the full ROAD dataset, merged and oversampled. This creates the results with the highest fidelity to the real-world scenario. On the contrary, the TAN, the DCNN and the improved DCNN version, together with the binary version of the LSTM-based model, perform binary classification for every different attack type, and they are trained and tested on each attack type's data which has been separately merged to resemble the setup of the HCRL Car-Hacking dataset and In-Vehicle Network Intrusion Detection Challenge dataset, in order to compare the results, so they eventually produce a different model per attack class.

Since the choice of performance evaluation metrics is crucial in well representing a model’s effectiveness for intrusion detection systems, we have opted for precision, recall and the F1 score, in order to correctly detect false positives and false negatives, while preserving a general value useful to evaluate the overall performance and therefore compare the models.

The experimental environment in which every model has been tested is on a system with a NVIDIA A100 GPU with 40GB of VRAM, 387.57GB of RAM, and two AMD EPYC 7302 16-Core Processors. The libraries to recreate the machine learning and deep learning models include PyTorch, TensorFlow and Scikit-Learn.

\subsection{Experimental Results and Discussion of the Models}

\subsubsection{LCCDE and ML approaches}
The performance of the LCCDE model, consisting of three gradient boosting classifiers (LightGBM, CatBoost and XGBoost), has been reproduced on the HCRL Car-Hacking dataset and we obtained very good and constant results on the In-Vehicle Network Intrusion Detection Challenge dataset. However, the model has not performed as well on the ROAD dataset, despite having analyzed both the ID and data field, with an average F1 score of 0.3212 and standard deviation of 0.1723. It is worth noting that this model could not detect at all two of the five masquerade attacks. A similar situation involves Random Forests which on average outperform slightly LightGBM, but had worse results than LCCDE. Furthermore, the LCCDE model was the one that required the most computation time due to the original implementation and impossibility of parallelization, requiring about 32 hours and 40 minutes of running time for the classification task on the ROAD dataset, while LightGBM has only taken about 6 minutes, XGBoost has taken about 10 minutes and CatBoost has taken about one hour and 27 minutes, which means that the vast majority of the computation time is due to the LCCDE algorithm. The code for the LCCDE model was made available by the authors \cite{lccde-code}.

Despite being more explainable and lightweight in terms of memory usage, these models were not able to reach satisfactory results on the ROAD dataset, therefore we can conclude that traditional machine learning approaches are not fit for this task.

\subsubsection{LSTM-based model}
The model showed an exceptional performance for binary classification on the ROAD dataset (Table \ref{tab:binary_results}), perfectly classifying almost every attack type except the fuzzing attack, where it still reached a very good F1 score of 0.9977. Unfortunately, a binary classification approach is not ideal for the purpose of an IDS, as it would require a different pre-trained model for every different attack type, making it particularly expensive and challenging to update. We can conclude that its real-world application is questionable.

For the multiclass classification on the ROAD dataset, the LSTM-based model was able to achieve, on average, the best performance compared to the other traditional machine learning models. Nonetheless, its performance is largely unsatisfying and the perfect recall and low precision for most of the attack types show a tendency to produce a large number of false positives. The model was able to obtain a good F1 score on the fuzzing attack thanks to the simplicity in detecting the data field set to the maximum value (\texttt{0xFFFFFFFFFFFFFF}) for every attack message. As for the other models, it reached particularly negative results on the Max Engine Coolant Temp Attacks for which it was unable to correctly classify any attack sample in the test set, likely due to their insufficient number (Table \ref{tab:road-num-logs}).

The opposite goes for the model's results on the Car-Hacking dataset and on the In-Vehicle Network Intrusion Detection Challenge dataset, where its average F1 score is 0.9778, while on the ROAD dataset is 0.5963, highlighting a great performance divergence between detecting simpler attacks and more realistic ones.

\subsubsection{DCNN} We have been able to reproduce the performance shown on the HCRL Car-Hacking dataset, which was built by the same authors of this model \cite{SONG2020100198}. The results were not as stable on the In-Vehicle Network Intrusion Detection Challenge dataset, having a higher average standard deviation of 0.05 on the F1 score metric, while the model has seen worse performance and dramatic fluctuations on the ROAD dataset. This has probably been caused by the fact that this dataset contains attacks acting also on the data field and they are much more difficult to detect because of their stealthiness, described on the Section~\ref{road-dataset-description}. Considering this, the DCNN has performed fairly well, sometimes producing less false positives than the TAN model, with a notable exception for the fuzzing attack. This attack is easily identifiable by the payload, always set on the maximum value, but this model only uses the ID field, that in this case has cycling IDs. The cycling pattern is hardly detectable from the input frame containing 29 consecutive IDs, which explains the unreliability of the model.

As mentioned in Subsection \ref{subsection:DCNN}, the original frame generation of the DCNN is using a sliding window of size 29 with a step of 29, which means that some attack patterns shown in ID bits could be cut into two different frames. For this reason we created a different version of frame generation, employing a step equal to one, and we tested it on the ROAD dataset, labelling it as DCNNv2. As Table \ref{tab:binary_results} shows, thanks to this small modification, the model could reliably and constantly improve its F1 score, going from an average of 0.6514 to an average of 0.7468, and it was able to beat the F1 score of the TAN for the majority of the attack types. Nonetheless, as for the LSTM-based model and the TAN, this approach is hardly applicable in a real-world scenario, as it would require multiple models running in parallel to detect only predefined attacks, other than a non negligible overhead for the frame generation.

\subsubsection{TAN}
The TAN was able to maintain a good performance for most of the attacks, and it reached very good F1 scores for most of the attack types except for the Max Engine Coolant Temp Attacks, which, as mentioned earlier, only have few attack samples available. Furthermore, its recall is always higher than its precision (as Table \ref{tab:binary_results} shows), even drastically in the case of the variations of the Max Engine Coolant Temp Attacks, which means that it produces more false positives than false negatives.

Like we previously mentioned, this model performs binary classification on sequences of CAN IDs, so it cannot distinguish the attack class detected and it was tested on every attack type singularly.

\begin{table}
\centering
\caption{ Model performance comparison for each class in the HCRL Car Hacking dataset.}
\begin{tabular}{|p{1.8cm}|c|r|r|r|}
\hline
\textbf{Attacks} & \textbf{Model} &  \bfseries\makecell{Precision} & \bfseries\makecell{Recall} & \bfseries\makecell{F1 score}\\
\hline
\multirow{4}{1.8cm}{DoS Attack} 
&RF     &\textbf{1.0000}&\textbf{1.0000}&\textbf{1.0000}\\
&LighGBM&\textbf{1.0000}&\textbf{1.0000}&\textbf{1.0000}\\
&LCCDE	&\textbf{1.0000}&\textbf{1.0000}&\textbf{1.0000}\\
&DCNN	&\textbf{1.0000}&0.9989&0.9995\\
&TAN	&\textbf{1.0000} &0.9961& 0.9980\\
&LSTM 	&\textbf{1.0000}&\textbf{1.0000}&\textbf{1.0000}\\

\hline
\multirow{4}{1.8cm}{Fuzzy Attack} 
&RF     &\textbf{1.0000}&\textbf{1.0000}&\textbf{1.0000}\\
&LighGBM&\textbf{1.0000}&\textbf{1.0000}&\textbf{1.0000}\\
&LCCDE	&\textbf{1.0000}&\textbf{1.0000}&\textbf{1.0000}\\
&DCNN	&0.9995&0.9965&0.9980\\
&TAN	&\textbf{1.0000}&0.9999&0.9999\\
&LSTM 	&0.9590&0.8802&0.9179\\

\hline
\multirow{4}{1.8cm}{Gear Spoofing Attack} 
&RF  &\textbf{1.0000}&\textbf{1.0000}&\textbf{1.0000}\\
&LighGBM&\textbf{1.0000}&\textbf{1.0000}&\textbf{1.0000}\\
&LCCDE  &\textbf{1.0000}&\textbf{1.0000}&\textbf{1.0000}\\
&DCNN   &0.9999&0.9989&0.9994\\
&TAN    &0.9989&0.9926&0.9957\\
&LSTM   &0.9592&\textbf{1.0000}&0.9792\\

\hline
\multirow{4}{1.8cm}{RPM Spoofing Attack} 
&RF &\textbf{1.0000}&\textbf{1.0000}&\textbf{1.0000}\\
&LighGBM  &\textbf{1.0000}&\textbf{1.0000}&\textbf{1.0000}\\
&LCCDE	&\textbf{1.0000}&\textbf{1.0000}&\textbf{1.0000}\\
&DCNN	&0.9999&0.9994&0.9996\\
&TAN	&0.9996&0.9967&0.9982\\
&LSTM 	&0.9998&\textbf{1.0000}&\textbf{1.0000}\\

\hline

\end{tabular}
\end{table}

\begin{table}
\centering
\caption{Performance evaluation of models on the In-Vehicle Network intrusion detection Challenge dataset.}
\begin{tabular}{|p{1.4cm}|c|r|r|r|}
\hline
\textbf{Vehicle} & \textbf{Model} & \bfseries\makecell{Precision} & \bfseries\makecell{Recall} & \bfseries\makecell{F1 score}\\
\hline
 \multirow{4}{1.7cm}{Chevrolet Spark}
&RF	        &99.9984&\textbf{99.9984}&\textbf{99.9984}\\
&LightGBM	&99.9952&99.9952&99.9952\\
&LCCDE      &99.9991&99.9167&99.9578\\
&DCNN       &85.5856&97.9603&90.1321\\
&TAN        &\textbf{100.0000}&99.7443&99.8719\\
&LSTM       &98.5301&99.9083&99.2144\\

\hline
\multirow{4}{1.7cm}{Hyundai Sonata} 
&RF	        &\textbf{100.0000}&\textbf{100.0000}&\textbf{100.0000}\\
&LightGBM	&99.9936&99.9936&99.9936\\
&LCCDE      &\textbf{100.0000}&\textbf{100.0000}&\textbf{100.0000}\\
&DCNN       &\textbf{100.0000}&99.1681&99.5816\\
&TAN        &99.9287&99.4970&99.8010\\
&LSTM       &\textbf{100.0000}&99.9265&99.9632\\

\hline
\multirow{4}{1.7cm}{Kia Soul} 
&RF	        &\textbf{99.9974}&\textbf{99.9974}&\textbf{99.9974}\\
&LightGBM	&99.9965&99.9965&99.9965\\
&LCCDE	    &99.9226&99.9828&99.9526\\
&DCNN       &99.3859&99.7984&99.5909\\
&TAN        &99.7438&99.6071&99.6754\\
&LSTM       &99.9877&95.8677&97.8844\\

\hline
\end{tabular}
\end{table}

\begin{table}
\centering
\caption{Model performance comparison for multiclass classification on the ROAD data.}
\begin{tabular}{|p{1.9cm}|c|r|r|r|r|}
\hline
\textbf{Attacks} & \textbf{Model} & \bfseries\makecell{Precision} & \bfseries\makecell{Recall} & \bfseries\makecell{F1 score}\\
\hline
 \multirow{4}{1.9cm}{Correlated Signal Fabrication Attack}
&RF         &\textbf{0.5068}&\textbf{1.0000}&0.6727\\
&LightGBM   &0.0000&0.0000&0.0000\\
&LCCDE      &0.5014&\textbf{1.0000}&\textbf{0.6766}\\
&LSTM       &0.5018&\textbf{1.0000}&0.6683\\
\hline
\multirow{4}{1.9cm}{Correlated Signal Masquerade Attack} 
&RF         &0.0000&0.0000&0.0000\\
&LightGBM   &\textbf{0.2725}&\textbf{1.0000}&\textbf{0.4213}\\
&LCCDE      &0.0000&0.0000&0.0000\\
&LSTM       &0.0000&0.0000&0.0000\\

\hline
\multirow{4}{1.9cm}{Fuzzing Attack} 
&RF         &\textbf{1.0000}&\textbf{1.0000}&\textbf{1.0000}\\
&LightGBM   &0.0134&0.6236&0.0311\\
&LCCDE      &0.2674&\textbf{1.0000}&0.4224\\
&LSTM       &0.9951&\textbf{1.0000}&0.9975\\

\hline
\multirow{4}{1.9cm}{Max Engine Coolant Temp Fabrication Attack}
&RF         &\textbf{0.6667}&\textbf{1.0000}&\textbf{0.8000}\\
&LightGBM   &0.0000&0.0000&0.0000\\
&LCCDE      &0.2500&\textbf{1.0000}&0.4000\\
&LSTM       &0.0000&0.0000&0.0000\\

\hline
\multirow{4}{1.9cm}{Max Engine Coolant Temp Masquerade Attack} 
&RF         &0.0000&0.0000&0.0000\\
&LightGBM   &0.0000&0.0000&0.0000\\
&LCCDE      &0.0000&0.0000&0.0000\\
&LSTM       &\textbf{0.7500}&\textbf{1.0000}&\textbf{0.8571}\\

\hline
\multirow{4}{1.9cm}{Max Speedometer Fabrication Attack} 
&RF         &0.1257&0.1204&0.1229\\
&LightGBM   &0.2944&\textbf{0.4512}&\textbf{0.3536}\\
&LCCDE      &0.3713&0.2384&0.2822\\
&LSTM       &\textbf{0.4335}&0.0586&0.1032\\

\hline
\multirow{4}{1.9cm}{Max Speedometer Masquerade Attack}
&RF         &0.1408&0.1469&0.1438\\
&LightGBM   &0.1715&0.1647&0.1693\\
&LCCDE      &0.3612&0.3612&0.3612\\
&LSTM       &\textbf{0.4976}&\textbf{0.9241}&\textbf{0.6469}\\

\hline
\multirow{4}{1.9cm}{Reverse Light Off Fabrication Attack}
&RF         &0.2514&0.2421&0.2467\\
&LightGBM   &0.0646&0.3173&0.1125\\
&LCCDE      &\textbf{0.3655}&0.3836&\textbf{0.3712}\\
&LSTM       &0.2432&\textbf{0.4684}&0.3201\\

\hline
\multirow{4}{1.9cm}{Reverse Light Off Masquerade Attack}
&RF         &0.2719&0.2846&0.2781\\
&LightGBM   &0.2511&0.3453&0.2946\\
&LCCDE      &\textbf{0.3612}&0.3334&0.3454\\
&LSTM       &0.2964&\textbf{0.4888}&\textbf{0.3690}\\

\hline
\multirow{4}{1.9cm}{Reverse Light On Fabrication Attack} 
&RF         &0.1567&0.1532&0.1550\\
&LightGBM   &0.2302&0.4988&0.3133\\
&LCCDE      &0.3221&0.3423&0.3384\\
&LSTM       &\textbf{0.4310}&\textbf{0.8239}&\textbf{0.5659}\\

\hline
\multirow{4}{1.9cm}{Reverse Light On Masquerade Attack}
&RF         &0.1702&0.1741&0.1721\\
&LightGBM   &0.0741&0.1889&0.1010\\
&LCCDE      &0.3222&\textbf{0.2993}&\textbf{0.3144}\\
&LSTM       &\textbf{0.3951}&0.1404&0.2071\\

\hline
\end{tabular}
\label{tab:multiclass_results}
\end{table}

\begin{table}
\centering
\caption{Model performance comparison for binary classification on the ROAD data.}
\begin{tabular}{|p{1.9cm}|c|r|r|r|r|}
\hline
\textbf{Attacks} & \textbf{Model} & \bfseries\makecell{Precision} & \bfseries\makecell{Recall} & \bfseries\makecell{F1 score}\\
\hline
 \multirow{4}{1.9cm}{Correlated Signal Fabrication Attack}
&DCNN	&0.9919&0.9388&0.9644\\
&DCNNv2	&0.9981&0.9889&0.9934\\
&TAN	&0.9504&\textbf{1.0000}&0.9746\\
&LSTM 	&\textbf{1.0000}&\textbf{1.0000}&\textbf{1.0000}\\

\hline
\multirow{4}{1.9cm}{Correlated Signal Masquerade Attack} 
&DCNN	&0.9947&0.6999&0.8130\\
&DCNNv2	&0.9997&0.9194&0.9578\\
&TAN	&0.9437&\textbf{1.0000}&0.9713\\
&LSTM 	&\textbf{1.0000}&\textbf{1.0000}&\textbf{1.0000}\\

\hline
\multirow{4}{1.9cm}{Fuzzing Attack} 
&DCNN	&0.2939&0.3104&0.2430\\
&DCNNv2	&0.2929&0.1264&0.1764\\
&TAN	&\textbf{0.9960}&0.9994&0.9976\\
&LSTM 	&0.9955&\textbf{1.0000}&\textbf{0.9977}\\

\hline
\multirow{4}{1.9cm}{Max Engine Coolant Temp Fabrication Attack}
&DCNN	&0.0836&0.6571&0.1323\\
&DCNNv2	&0.8163&0.0744&0.1230\\
&TAN    &0.5334&\textbf{1.0000}&0.6957\\
&LSTM 	&\textbf{1.0000}&\textbf{1.0000}&\textbf{1.0000}\\

\hline
\multirow{4}{1.9cm}{Max Engine Coolant Temp Masquerade Attack} 
&DCNN	&0.1161&0.4381&0.1369\\
&DCNNv2	&0.6510&0.1044&0.1700\\
&TAN	&0.0098&0.9535&0.0194\\
&LSTM 	&\textbf{1.0000}&\textbf{1.0000}&\textbf{1.0000}\\

\hline
\multirow{4}{1.9cm}{Max Speedometer Fabrication Attack} 
&DCNN	&0.9457&0.5039&0.6511\\
&DCNNv2	&0.9617&0.8543&0.9049\\
&TAN	&0.9884&\textbf{1.0000}&0.9941\\
&LSTM 	&\textbf{1.0000}&\textbf{1.0000}&\textbf{1.0000}\\

\hline
\multirow{4}{1.9cm}{Max Speedometer Masquerade Attack}
&DCNN	&0.9893&0.4629&0.6058\\
&DCNNv2	&0.9998&0.8880&0.9404\\
&TAN	&0.9881&0.9986&0.9933\\
&LSTM 	&\textbf{1.0000}&\textbf{1.0000}&\textbf{1.0000}\\

\hline
\multirow{4}{1.9cm}{Reverse Light Off Fabrication Attack}
&DCNN	&0.7945&0.9733&0.8710\\
&DCNNv2	&0.9533& 0.9992&0.9756\\
&TAN	&0.9753&0.9884&0.9818\\
&LSTM 	&\textbf{1.0000}&\textbf{1.0000}&\textbf{1.0000}\\

\hline
\multirow{4}{1.9cm}{Reverse Light Off Masquerade Attack}
&DCNN	&0.9852&0.8000&0.8780\\
&DCNNv2	&0.9981&0.9953&0.9967\\
&TAN    &0.9641&0.9883&0.9761\\
&LSTM 	&\textbf{1.0000}&\textbf{1.0000}&\textbf{1.0000}\\

\hline
\multirow{4}{1.9cm}{Reverse Light On Fabrication Attack}
&DCNN &0.8871&0.9644&0.9232\\
&DCNNv2	&0.9600&0.9951&0.9772\\
&TAN &0.9600&0.9743&0.9670\\
&LSTM 	&\textbf{1.0000}&\textbf{1.0000}&\textbf{1.0000}\\

\hline
\multirow{4}{1.9cm}{Reverse Light On Masquerade Attack}
&DCNN   &0.9993&0.9069&0.9467\\
&DCNNv2	&0.9998&0.9990&0.9994\\
&TAN    &0.9593&0.9744&0.9668\\
&LSTM 	&\textbf{1.0000}&\textbf{1.0000}&\textbf{1.0000}\\

\hline
\end{tabular}
\label{tab:binary_results}
\end{table}

\section{Conclusion}\label{sec:conclusion}

After an accurate analysis of different models on the novel ROAD dataset and two datasets established in previous literature, the HCRL Car-Hacking dataset and the In-Vehicle Network Intrusion Detection Challenge dataset, we encountered a sensible disparity in the results. Every model performing multiclass classification that had shown almost perfect predictions on the former datasets, demonstrated to be highly ineffective on the latter, rarely reaching satisfactory results. On the contrary, the models performing binary classification had kept good precision, recall and F1 score for most of the attack types. However, this is not due to the intrinsic difficulty of multiclass classification of multivariate data, but to the setup in which the latter have been usually tested, which involves binary classification performed on different datasets, each containing one isolated attack class. This case is scarcely indicative of a real-world scenario because, to detect each class, it would require multiple models being deployed at the same time, each trained on one predefined attack, therefore increasing the resources required as well as compromising their ability of detecting new attack types as well. Thus, this setup produces exceptionally good metrics that unfortunately do not translate well with the models' practical performance. Furthermore, the authors of these models have not taken into consideration their complexity in terms of memory and processing time, which would be a fundamental aspect of in-vehicle detection, due to the limited computational resources available.

Given the stealthier attacks of the ROAD dataset which better represent the real-world case, this dataset should be considered as more fit to represent a realistic scenario, therefore it should be included in the test benches on which new IDSs would be benchmarked, and new datasets should be built aiming in this direction.

However, as shown by the number of samples on Table~\ref{tab:road-num-logs}, the ROAD dataset is not without flaw, presenting a number of samples too small for deep learning approaches and a heavy class imbalance, particularly for the Max Engine Coolant Temp Fabrication Attack and the corresponding masquerade attack. Indeed, every model produced their worst results on these two classes, supporting the hypothesis that the number of attack samples is insufficient for this task.

\begin{acks}
This paper is supported by the European Union’s HORIZON Research and Innovation Programme under grant agreement No 101120657, project ENFIELD (European Lighthouse to Manifest Trustworthy and Green AI). The authors also appreciate the support from the University of Bologna which granted Lorenzo Guerra a scholarship for the preparation of his Master's thesis abroad.
\end{acks}

\bibliographystyle{ACM-Reference-Format}
\bibliography{bibliography}

\end{document}